\documentclass{svjour3}                     
\smartqed  
\usepackage{graphicx}
\usepackage{mathptmx}      

\usepackage[latin1]{inputenc}
\usepackage{amsfonts}
\usepackage{mathrsfs}
\usepackage{amsmath}
\usepackage{amssymb}
\usepackage[T1]{fontenc}
\usepackage{graphicx}
\newcommand{\ud}{{ \, \mathrm{d}}}
\newcommand{\nd}{\stackrel{def}{=}}

\def \qed{\penalty-20\null\hfill$\square$\par\medbreak}

\def \Tstar {T^*}

\newtheorem{Theorem}{Theorem}[section]
\newtheorem{Definition}[Theorem]{Definition}
\newtheorem{Proposition}[Theorem]{Proposition}
\newtheorem{Lemma}[Theorem]{Lemma}

\newtheorem{Hypothesis}[Theorem]{Hypothesis}
\newtheorem{Remark}[Theorem]{Remark}

\newtheorem{Notation}[Theorem]{Notation}

\begin{document}

\title{An Hilbert space approach for a class of arbitrage free implied volatilities models\thanks{Work supported by the ARC Discovery project DP0558539.}}


\author{A. Brace \and G. Fabbri \and B. Goldys}


\institute{A. Brace \at National Australia Bank and Financial Mathematical Modelling and Analysis. E-mail: abrace@ozemail.com.au
\and
G. Fabbri \at
School of Mathematics and Statistics, UNSW, Sydney. \email{gieffe79@gmail.com}           
           \and
           B. Goldys \at
              School of Mathematics and Statistics, UNSW, Sydney. E-Mail: beng@maths.unsw.edu.au.
}

\date{Received: date / Accepted: date}

\maketitle

\begin{abstract}
We present an Hilbert space formulation for a set of implied volatility models introduced in \cite{BraceGoldys01} in which the authors studied conditions for a family of European call options, varying the maturing time and the strike price $T$ an $K$, to be arbitrage free. The arbitrage free conditions give a system of stochastic PDEs for the evolution of the implied volatility surface ${\hat\sigma}_t(T,K)$. We will focus on the family obtained fixing a strike $K$ and varying $T$. In order to give conditions to prove an existence-and-uniqueness result for the solution of the system it is here expressed in terms of the square root of the \textit{forward implied volatility} and rewritten in an Hilbert space setting. The existence and the uniqueness for the (arbitrage free) evolution of the forward implied volatility, and then of the the implied volatility, among a class of models, are proved. Specific examples are also given.
\hskip 0.0cm

\end{abstract}

\section*{Introduction}
The main aim of the paper is to prove an existence-and-uniqueness result, to study properties of the solution and to give some examples for the implied volatility model presented in \cite{BraceGoldys01}: in such a seminal work the authors presented a set of conditions, written as a system of SPDEs, for the market (described below) to be arbitrage free. Here we prove that, indeed, under a suitable set of conditions, such a system of SPDEs admits a (unique) solution.

In other words the results we give allow to identify a class of (non-trivial, arbitrage free) evolutions of the implied volatility  starting from some the initial (market-given) surface.

Many aspects of implied volatility models have been diffusely studied and the reader is referred to \cite{MusielaRutkowski05}, Chapter 7 for a review.

\subsection*{The setting of the model and some results from \cite{BraceGoldys01}}
Consider $W^{(i)}_t$, for $i\in \{ 1, .. , m \}$ and $t\geq 0$, $m$ independent real Brownian Motions  on the probability space $(\Omega, \mathcal{F}, \mathbb{P})$. We call $\mathcal{F}_t$ the induced filtration. We consider a fixed $T^*>0$ and a market in which a bond (with interest rate equal to zero), a stock $S_t$ and a family of European call options $O_t(K,T)$ for $t\geq 0$, $T\in (t,t+T^*]$, and $K>0$ are liquidly traded. So at every time $t$ we consider the call options expiring in the interval $(t,t+T^*]$ for a fixed $T^*$. Without losing in generality (changing if necessary the Brownian motions and the measure $\mathbb{P}$) we can assume that the price of the stock $S_t$ depends only on the first BM, that $S_t$ is martingale and evolves following the SDE
\begin{equation}
\label{eq:intro-evol-S}
\ud S_t = S_t \theta_t \ud W_t^{(1)}
\end{equation}
for some one-dimensional process $\theta_t$. 
The Black and Scholes price for $O_t(T,K)$ is  of course 
\begin{equation}
\label{eq:BSprice}
C_t(S_t,\sigma,K,T)=S_tN(d_{1}(S_t,\sigma,K,T))-KN(d_{2}(S_t,\sigma,K,T))
\end{equation}
where $N$ is the cumulative distribution of the normal distribution and, 
\[
d_{1}(S_t,\sigma,K,T)=\frac{\ln\frac{S_{t}}{K}}{\sigma\sqrt{T-t}} + \frac{1}{2}\sigma \sqrt{(T-t)}, \qquad d_{2}(S_t,\sigma,K,T)=\frac{\ln\frac{S_{t}}{K}}{\sigma\sqrt{T-t}} - \frac{1}{2}\sigma \sqrt{(T-t)}.
\]
The implied volatility paradigm consists, as well known, in inverting (\ref{eq:BSprice}) obtaining (and defining) the ``(Black-Scholes) implied volatility'' $\hat\sigma_t(T,K)$ as a function of $C_t$ (and $K$, $T$, $S_t$).
So, once we have modeled the evolution of the implied volatility, thanks to its definition, we can use (\ref{eq:BSprice}) to find the evolution (varying the time $t$) of the prices of the options $O_t(T,K)$ and we can wonder if the evolution of the market so obtained is arbitrage free
namely, if the processes $C_t(T,K):= S_tN(d_{1}(S_t,{\hat\sigma}_t,K,T))-KN(d_{2}(S_t,{\hat\sigma}_t,K,T))$ and $S_t$ have an equivalent common (varying $T$ and $K$) local martingale measure.

In \cite{BraceGoldys01} the authors prove that, if we assume the  implied volatility to follow a SDE of the form\footnote{Where $m_t$ and $v_t$ are respectively a one-dimensional and a $m$-dimensional process and they can depend explicitly, as we will assume when we give some sufficient conditions to prove the existence of the solution, on $T$, $K$, $S_t$, ${\hat\sigma}_t$ and $\theta_t$. $v_t(T,K)^*$ is the adjoint of the vector $v_t(T,K)$ so that $v_t(T,K)^* \ud W_t= \left\langle v_t(T,K), \ud W_t \right \rangle$ ($\left\langle \cdot , \cdot \right\rangle$ represents the scalar product in $\mathbb{R}^m$).}
\[
\ud {\hat\sigma}_t(T,K) = m_t(T,K) \ud t + v_t(T,K)^* \ud W_t,
\]
the arbitrage-free conditions for the market can be expressed (we do not write the dependence of ${\hat\sigma}_t$, and $u_t: =v_t/{\hat\sigma}_t$ on $T$ and $K$ in the second equation) as
\begin{equation}
\label{eq:BLforulation-sigma}
\left \{
\begin{array}{rcl}
\multicolumn{3}{l}{\ud S_t = S_t\theta_t \ud W_t^{(1)}}\\
\ud {\hat\sigma}_t & = & \frac{1}{2{\hat\sigma}_t(T-t)}\left ( {\hat\sigma}_t^2 - \left | \theta_t \ell + u_t \ln \frac{K}{S_t} \right |^2 \right ) \ud t + \\
& & + \left ( \frac{1}{8} {\hat\sigma}_t^3(T-t) |u_t|^2 - \frac{1}{2}{\hat\sigma}_t \theta_t u_t^{(1)} \right ) \ud t + {\hat\sigma}_t u_t^* \ud W_t\\
\multicolumn{3}{l}{{\hat\sigma}_0({T,K}) \qquad\qquad \qquad\qquad\qquad \text{initial condition}}\\
\multicolumn{3}{l}{{\hat\sigma}_t(T,K) = \left | \theta_t\ell + u_t \ln \frac{K}{S_t} \right | \qquad\;\;\; \text{feedback condition}}.
\end{array}
\right .
\end{equation}
where we called $\ell$ the vector of $R^m$ given by $(1,0,0,...,0)$, $|\cdot|$ is the norm in $\mathbb{R}^m$ and the $m$-dimensional process $u_t=v_t/{\hat\sigma}_t$. They also prove that such a system of SPDEs can be rewritten using the variable\footnote{See also \cite{BraceGoldys02}.} $\xi_t(T,K) = (T-t){\hat\sigma}^2_t(T,K)$ obtaining
\begin{equation}
\label{eq:BLforulation}
\left \{
\begin{array}{rcl}
\multicolumn{3}{l}{\ud S_t = S_t\theta_t \ud W_t^{(1)}}\\
\ud \xi_t & = & \xi_t \left ( \left ( 1+ \frac{1}{4} \xi_t \right ) |u_t|^2 - \theta_t u_t^{(1)}  \right ) \ud t - \left ( \theta_t + u_t^{(1)} \ln \left( \frac{K}{S_t} \right) \right )^2 \ud t -\\ 
& & - \sum_{i=2}^{m} \left ( u_t^{(i)} \right )^2 \ln^2 \left( \frac{K}{S_t} \right ) \ud t +2 \xi_t u_t^* \ud W_t\\
\multicolumn{3}{l}{\xi_0^{T} \qquad\qquad\qquad\qquad\qquad\qquad\qquad\qquad\qquad\qquad\qquad\qquad\qquad\;\;\;\,\text{ initial condition}}\\
\multicolumn{3}{l}{\left . \partial_T \xi_t (T,K) \right |_{T=t} = \left ( \theta_t + u_t^{(1)} \ln \left( \frac{K}{S_t} \right) \right )^2 + \sum_{i=2}^{m} \left ( u_t^{(i)} \right )^2 \ln^2 \left( \frac{K}{S_t} \right ) \;\;\; \text{feedback condition}}
\end{array}
\right .
\end{equation}
where we used $u_t^{(i)}$ for the $i$-th component of $u_t$.

The feedback condition is obtained in \cite{BraceGoldys01} in order to avoid the phenomenon (already observed in \cite{Shombucher99}, Section 3(a), see also (\cite{berestyckibuscaflorent02} and \cite{berestyckibuscaflorent04})) of the ``bubble'' of the drift for $t\to T$. Such a condition, it will be clearer in the following, adds a certain number of difficulties in the study of the problem.

In \cite{BraceGoldys01} the author does not prove an existence result for equation (\ref{eq:BLforulation-sigma}) or (\ref{eq:BLforulation}) but they prove that such conditions are equivalent to the  market being arbitrage-free. So, if we can find some sets of $u^{(i)}_t$ and $\theta_t$ of stochastic processes such that equations (\ref{eq:BLforulation}, \ref{eq:intro-evol-S}) admit a positive solution $(\xi_t, S_t)$ (or, that is the same, (\ref{eq:BLforulation-sigma}, \ref{eq:intro-evol-S}) admit a positive solution $({\hat\sigma}_t,S_t)$), the evolution of the market is arbitrage free.

In the present work we study a ``reduced'' problem: indeed we consider a fixed $K$ and we study the existence and uniqueness for the system of SPDEs (\ref{eq:BLforulation}) varying $T$. We continue in the introduction to write the equations for the the general problem and we will fix a $K$ in Section \ref{sec:theproblem} (starting from equation (\ref{eqstateinfdim})). For the general case we would need the ``compatibility conditions'' described in Section \ref{sec:varyingK} to be satisfied.

\subsection*{Forward implied volatility and formal derivation of the state equation}
We want to describe the system using the square root of the \textit{forward implied volatility} introduced in \cite{Shombucher99}. We define $X_t$, formally, as
\begin{equation}
\label{eq:defimpliedvolatility}
{X}_t(x,K)= \frac{\partial }{\partial x} \left ( x {\hat\sigma}^2_t(x,K) \right ) = \frac{\partial }{\partial x} \left ( \xi_t(t+x,K) \right ).
\end{equation}
The idea of use such a variable in the implied volatility models was introduced for the first time, as far as we know, in \cite{SchweizerWissel06}. In the works \cite{SchweizerWissel06,SchweizerWissel2007-strike} the authors use different techniques to deal with problems strictly related to the our. They use some results about strong solutions for functional SDE proven in \cite{Wissel07} (see also \cite{Protter05}) to study the case of the family for a fixed strike $K$ (varying the maturing time $T$) in \cite{SchweizerWissel06} and the family for fixed $T$ (varying the strike) in \cite{SchweizerWissel2007-strike}.

The main novelty with respect to the results obtained in \cite{SchweizerWissel06} concerns the mathematical techniques used, but the Hilbert space approach used in the present works allows also to avoid a couple of additional ``technical conditions'' required\footnote{Actually to compare the Hypotheses needed in the two different setting is not very easy.} in \cite{SchweizerWissel06} and to use an analogous of the Musiela parameterization (see \cite{Musiela93}) for the HJM interest rate model. So we can consider at every time $t$ the family of call option (for a fixed $K$) for all the expiration times $T\in [t,t+T^*]$ for a fixed $T^*$.

\medskip

\noindent
From the second equation of (\ref{eq:BLforulation}), using Ito-Venttsel formula (see for example \cite{Rozovskii73}), we have, formally:
\begin{multline}
\label{eq:evolutionforKinRm}
\ud { X}_t(x,K)= \left [ \partial_x { X}_t(x,K) + \left ( \int_0^x { X}_t(r,K) \ud r \right ) \Bigg (\frac{1}{2} { X}_t(x,K) |u_t|^2 + \right .\\
+\frac{1}{2} \left\langle u_t, \partial_x u_t \right\rangle \int_0^x { X}_t(r,K) \ud r + 2 \left\langle u_t, \partial_x u_t \right\rangle - \theta_t \partial_x u^{(1)}_t \Bigg ) +\\
+ { X}_t(x,K) \left ( |u_t|^2 - \theta_t u_t^{(1)} \right ) - \\ 
\left . - 2 \left\langle \theta_t \ell + u_t \ln\left( \frac{K}{S_t} \right ), (\partial_x u_t) \ln\left( \frac{K}{S_t} \right ) \right\rangle \right ] \ud t+\\
+\left [ 2 { X}_t(x,K) u^*_t + 2 (\partial_x u^*_t) \int_0^x { X}_t(r,K) \ud r \right ] \ud W(t)
\end{multline}
where $\left\langle \cdot , \cdot \right\rangle$ is the scalar product in $\mathbb{R}^m$ and $\ell$ as above.

So here we formally defined $X_t$ as derivative of $\xi_t$ and we formally obtained the differential equation that describes the evolution of the square root of the forward implied volatility $X_t$ from the equation for $\xi_t$. Such a differentiation is only formal and this way to approach the problem (the most natural way from the point of view of the model) is not mathematically rigorous. For this reason our approach will be ``reversed'', we will describe it more precisely in the next paragraph.

\subsection*{The Hilbert space setting and the rigorous approach to the problem: the case of a fixed $K$}
We treat the problem using a Hilbert space formulation. A similar approach was used for example in \cite{GoldysMusiela01} for the HJM interest rate model (see also \cite{Filipovic01}, \cite{Cont00} and \cite{RingerTehranchi06}).

We consider a fixed $T^*>0$ and the Hilbert space $\mathcal{H}:=H^1(0,T^*)$ (the Sobolev space of index $1$).

\begin{Notation}
We use the notation $f[x]$ to denote the evaluation of an element $f$ of $H^1(0,T^*)$ (or of $L^2(0,T^*)$) at the point $x\in [0,T^*]$.
\end{Notation}
We want to describe $X_t(x,K)$ as an element of $H^1(0,T^*)$. So we introduce $\tilde X_t(K)$ defined as
\[
\tilde X_t(K)[x] \nd X_t(x,K).
\]
Of course, given an arbitrary function $X_t(x,K)$ the function $\tilde X_t(K)$ will not necessary belongs to $H^1(0,T^*)$, but we will see that (under suitable conditions on the functions $u^{(i)}$) if the initial 
$\tilde X_0(K)$ is in $H^1(0,T^*)$, its evolution remains in $H^1(0,T^*)$. With an abuse of notation we will call $\tilde X_t(K)$ simply $X_t(K)$.

\noindent
We call $I$ the continuous linear application
\[
\left \{
\begin{array}{ll}
I\colon \mathcal{H}\to \mathcal{H}, \; \; \; f\mapsto I(f)\\
I(f)[x]= \int_0^x f[s] \ud s
\end{array}
\right .
\]
and $A$ the generator of the $C_0$ semigroup $T(t)$ on $\mathcal{H}$ defined, for $t\geq 0$, as
\begin{equation}
\label{eq:defsemigroup}
(T(t))f[x]=\left \{
\begin{array}{ll}
f[x+t] & if\; x+t\leq T^*\\
f[T^*] & if\; x+t > T^*
\end{array}
\right .
\end{equation}
so that
\[
\left \{
\begin{array}{ll}
D(A) = \left \{ f\in H^2 \; : \; f[T^*]=\frac{\ud}{\ud t}f[T^*]=0 \right \}\\
A(f) = \frac{\ud}{\ud t}f[\cdot]
\end{array}
\right .
\]

In order to introduce some assumptions to guarantee the existence of the solution we assume that $u_t^{(i)}$ depends directly on $X_t(K)$, $K$ and $S_t$ and so we write $u_t{(i)}(K,S_t,X_t(K))[x]$ (since $u_t(K,S_t,X_t(K))$ will be an $H^1(0,T^*;\mathbb{R}^m)$-valued process we write $u_t(K,S_t,X_t(K))[x]$ to mean the evaluation of $u_t(K,S_t,X_t(K))$ at the point\footnote{Note that $u_t(K,S_t,X_t(K))[x]$ depends not only on $X_t(K)[x]$ but on all $X_t(K)[\cdot]$.} $x\in[0,T^*]$, so $u_t(K,S_t,X_t(K))[x]\in\mathbb{R}^m$). We write $\partial_x u_t(K,S_t,X_t(K))$ for\footnote{So, since $u_t(K,S_t,X_t)\in H^1(0,T^*)$ we have that $\partial_x u_t(K,S_t,X_t(K))$ is in $L^2(0,T^*)$. Actually we will give conditions to ensure that $\partial_x u_t(K,S_t,X_t(K))$ belongs to $H^1(0,T^*)$.} $\partial_x u_t(K,S_t,X_t(K))[x]$. So we can write formally equation (\ref{eq:evolutionforKinRm}) in $\mathcal{H}$ as:
\begin{equation}
\label{eq:stataoriginale}
\left \{
\begin{array}{rl}
\ud { X}_t(K) \!\!\!\! & =  AX_t(K) \ud t + \left [ I(X_t)(K) \Bigg (\frac{1}{2} {X}_t(K) |u_t(K,S_t,X_t(K))|^2 + \right .\\
&+\frac{1}{2} \left\langle u_t(K,S_t,X_t(K)), \partial_x u_t(K,S_t,X_t(K)) \right\rangle I(X_t)(K) +\\ 
&+ 2 \left\langle u_t(K,S_t,X_t(K)), \partial_x u_t(K,S_t,X_t(K)) \right\rangle - \theta_t(K,S_t,X_t(K)) \partial_x u_t^{(1)}(K,S_t,X_t(K))\ell \Bigg ) +\\
&+ { X}_t(K) \left ( |u_t(K,S_t,X_t(K))|^2 - \theta_t(K,S_t,X_t(K)) u_t(K,S_t,X_t(K))^{(1)} \right ) - \\ 
&\left . - 2 \left\langle \theta_t(K,S_t,X_t(K))\ell + u_t(K,S_t,X_t(K)) \ln\left( \frac{K}{S_t} \right ), (\partial_x u_t(K,S_t,X_t(K))) \ln\left( \frac{K}{S_t} \right ) \right\rangle \right ] \ud t+\\
&+\left [ 2 { X}_t(K) u^*_t(K,S_t,X_t(K)) + 2 (\partial_x u^*_t(K,S_t,X_t(K))) I(X_t)(K) \right ] \ud W(t)\\
\ud S_t = &\theta_t(K,S_t,X_t(K)) S_t \ud W_t^{(1)}\\
 X_0 (x,K) \!\!\!\!\!\!\!\!\! &  \;\;\qquad\textbf{initial condition}
\end{array}
\right .
\end{equation}
where $\left\langle \cdot, \cdot \right\rangle$ is the scalar product in $\mathbb{R}^m$ and $\theta_t(K,S,X)$ is for the following function (obtained is sing the fourth equation of (\ref{eq:BLforulation})):
\begin{equation}
\label{eq:theta}
\theta_t(K,S,X)= \sqrt{X[0] - \sum_{j=2}^m \left ( u^{(j)}_t(K,S,X)[0]\right )^2 \ln^2\left ( \frac{K}{S_t} \right )} - u_t^{(1)}(K,S,X)[0] \ln\left ( \frac{K}{S} \right ).
\end{equation}

This completes the ``informal'' formulation of the problem, the rigorous approach (the one that we develop in the paper) is reversed. 
We consider a fixed $K>0$, so we consider the family $O_t(T,K)$ for a fixed $K$ and varying $T\in [t,t+T^*]$. We start studying equation (\ref{eq:stataoriginale}, \ref{eq:theta}) in $\mathcal{H}$ and we will introduce the variable $\xi$ and the implied volatility problem only later. This is a scheme of our approach:
\begin{itemize}
\item[1.] We start (Section \ref{sec:theproblem}) studying the equations (\ref{eq:stataoriginale}, \ref{eq:theta}). So in Section \ref{sec:theproblem} we will introduce equation (\ref{eqstateinfdim}), that is nothing but a more concise form for (\ref{eq:stataoriginale}), without claiming any connection with the equation (\ref{eq:BLforulation}). 
\item[2.] We study (Section \ref{sec:Xepsilon}) some properties, existence and uniqueness results for (\ref{eqstateinfdim}) and its approximation (\ref{eqstateepsilon}).
\item[3.] We introduce (Section \ref{sec:xi}) $\xi_t(T,K)$ as $\xi_t(T,K):= I(X_t)[T-t]$ and prove that indeed such a $\xi_t(T,K)$ satisfy, as we expect, equation (\ref{eq:BLforulation}). We will use this fact also to prove that both $\xi_t(T,K)$ and $X_t(T,K)$ remain positive.
\end{itemize}
\noindent
In Section \ref{sec:examples} we present two classes of examples that verify the Hypotheses we described in Section \ref{sec:theproblem} (Hypothesis \ref{hp:suu} and \ref{hp:sqrtpositive}): in the first is the volvol does not depend on $T$ while in the second (more interesting) the volvol (that is the general statements is a function of $X_t$) depends in fact on $\xi_t$ in a quite general way. Note that a volvol that depends on $X_t$ through $\xi_t$ is exactly what we need to
write the equation (\ref{eq:BLforulation}) without the presence of $X_t$, and this is exactly the existence result we expected.

\section{Formulation of the problem and assumptions}
\label{sec:theproblem}
We consider a final time $T_0$. Later we will need to require that $T_0\leq T^*$. So we assume from now $T_0=T^*$.

Consider a probability space $(\Omega,\mathcal{F},\mathbb{P})$. Let $\mathcal{F}_t$ be the $\mathbb{P}$-augmented filtration generated by an $m$-dimensional Brownian motion $W_t$ (of components $W_t^{(i)}$ for $i=1,..,m$) for $t\geq 0$.
Let $u$ be a function
\[
\left \{
\begin{array}{l}
u \colon [0,\Tstar]\times \mathbb{R}^+ \times \mathbb{R}^+  \times H^1(0,T^*) \to H^1 (0,T^*;\mathbb{R}^m) = (H^1 (0,T^*))^m\\
(t,K,S,X) \mapsto (u_t(K,S,X)^{(1)}[\cdot], ... , u_t(K,S,X)^{(m)}[\cdot]).
\end{array}
\right .
\]
For $(t,K,S,X)\in [0,\Tstar]\times \mathbb{R}^+ \times \mathbb{R}^+  \times H^1(0,T^*)$ we define 
\begin{equation}
\label{eq:bartheta}
\bar\theta_t(K,S,X):= \sqrt{\left | X[0] - \sum_{j=2}^m \left ( u^{(j)}_t(K,S,X)[0]\right )^2 \ln^2\left ( \frac{K}{S} \right ) \right |} - u_t^{(1)}(K,S,X)[0] \ln\left ( \frac{K}{S} \right ).
\end{equation}
We assume that
\begin{Hypothesis}
\label{hp:suu}
For all $i\in \{ 1,..,m\}$
\[
\left \{
\begin{array}{l}
u^{(i)}\colon [0,\Tstar] \times \mathbb{R}^+ \times \mathbb{R}\times H^1(0,T^*) \to H^1(0,T^*)\\
u^{(i)} \colon (t, K,S,X) \mapsto u_t(K,S,X)
\end{array}
\right .
\]
is measurable from $\Big ( ([0,\Tstar] \times \mathbb{R}^+ \times \mathbb{R}\times H^1) ,  \mathcal{B}([0,\Tstar] \times \mathbb{R}^+ \times \mathbb{R}\times H^1) \Big )$ into $\Big (H^1,\mathcal{B}(H^1) \Big )$ where $\mathcal{B}$ is the $\sigma$-algebra generated by the Borel sets. Moreover we assume that, for all $K>0$ there exists a $C>0$ such that for all $t>0$ and for all $(S,X)\in \mathbb{R}^+\times H^1(0,T^*)$ we have
\begin{itemize}
\item[(i)] $\left | u_t^{(i)}(K,S,X)[x] \right | \leq C \frac{1}{1+ \left | \ln \left ( S \right ) \right | + \left | \int_0^x X(K)[s] \ud s \right | + |\bar\theta_t(K,S,X) |}$ for all $x\in [0,T^*]$
\item[(ii)] $\partial_x u_t^{(i)}(K,S,X)$ is in $H^1(0,T^*)$
\item[(iii)] $u_t^{(i)}(K,S,X)$, $\partial_x u_t^{(i)}(K,S,X)$, $u_t^{(1)}(K,S,X) \bar\theta_t(K,S,X)$, $\partial_x u_t^{(1)}(K,S,X) \bar\theta_t(K,S,X)$ are locally Lipschitz (as functions in $H(0,T^*)$) in $(S,X)\in \mathbb{R}^+\times H^1$ uniformly in $t$
\item[(iv)] $\left | \partial_x u_t^{(1)}(K,S,X) \right | \leq \frac{C}{1+|\bar\theta_t(K,S,X)|}$ and $\left | \partial_x u_t^{(1)}(K,S,X) \bar\theta_t(K,S,X)\ln(S) \right | \leq C(1+|X|)$
\end{itemize}
\end{Hypothesis}

\begin{Remark}
Note that we could treat a more general case, using the same arguments, allowing the explicit dependence of $u_t^{(i)}$ on $\omega\in\Omega$. In such a case we would require $u_t^{(i)}$ to be an adapted process and for all $i\in \{ 1,..,m\}$
\[
\left \{
\begin{array}{l}
u^{(i)}\colon [0,\Tstar] \times \Omega \times \mathbb{R}^+ \times (\mathbb{R}\times H^1(0,T^*)) \to H^1(0,T^*)\\
u^{(i)} \colon (t,\omega, K,(S,X)) \mapsto u_t(K,S,X)(\omega)
\end{array}
\right .
\]
is measurable from 
\[
\bigg ( ([0,\Tstar]\times \Omega) \times \mathbb{R}^+ \times (\mathbb{R}\times H^1) , \mathcal{P}_{\Tstar} \times \mathcal{B}(\mathbb{R}^+) \times \mathcal{B}(\mathbb{R}\times H^1) \bigg )
\]
into $\bigg ( H^1 ,\mathcal{B}(H^1) \bigg )$ where $\mathcal{B}$ is the $\sigma$-algebra generated by the Borel sets and $\mathcal{P}_{\Tstar}$ is the $\sigma$-field on $([0,\Tstar]\times \Omega)$ generated by the sets of the form $[s,t]\times F$ with $0\leq s <t<\Tstar$ and $F\in \mathcal{F}_s$. In this setting we have to ask that we have to ask $(i)$ .. $(iv)$ to be satisfied uniformly in $\omega\in\Omega$.\qed
\end{Remark}

\noindent
In order to avoid the absolute value in the definition of $\bar\theta$ and then came to the original problem we would like now to impose the following condition (that is implied by (\ref{eq:BLforulation}):
\begin{equation}
\label{eq:conditionwewouldlike}
X_t(K)[0] - \sum_{j=2}^m \left ( u^{(j)}_t(K,S_t,X_t)[0]\right )^2 \ln^2\left ( \frac{K}{S_t} \right ) \geq 0,
\end{equation}
but of course such a condition can be imposed only if $X_t(K)[0]\geq 0$.
We ask the following (we will see that it is enough to have (\ref{eq:conditionwewouldlike}) along the trajectories of the system)
\begin{Hypothesis}
\label{hp:sqrtpositive}
\[
X[0] - \sum_{j=2}^m \left ( u^{(j)}_t(K,S,X)[0]\right )^2 \ln^2\left ( \frac{K}{S} \right ) \geq 0 \; \; \forall S>0, \; \forall X\in H^1(0,T^*) \; with \; X[0]\geq 0
\]
and 
\[
X[0] - \sum_{j=2}^m \left ( u^{(j)}_t(K,S,X)[0]\right )^2 \ln^2\left ( \frac{K}{S} \right ) =0 \;\; \Longleftrightarrow \;\; X[0] = 0.
\]
\end{Hypothesis}
We impose that the initial data are strictly positive, this is a realistic assumption from the point of view of the model, note that in \cite{SchweizerWissel06} the authors argue (Proposition 2.1) that the negativity of the square root of the forward implied volatility causes elementary arbitrage opportunities:
\begin{Hypothesis}
\label{hp:X0positivo}
For every $K>0$ we choose the initial datum $(s_0,x_0)\in(\mathbb{R}\times H^1(0,T^*))$ with $s_0>0$ and $x_0>0$. This means, since $x_0>0$ is in $H^1$ and then it is continuous, that for every $K>0$ there exists a $c>0$ such that $x_0(K)[x]>0$ for all $x\in [0,T^*]$.
\end{Hypothesis}

\noindent
We define the functions
\begin{equation}
\begin{array}{l}
F\colon [0,\Tstar] \times \mathbb{R}^+ \times (\mathbb{R}\times H^1) \to H^1\\
B\colon [0,\Tstar] \times \mathbb{R}^+ \times (\mathbb{R}\times H^1) \to (H^1(0,T^*))^m = H^1(0,T^* ; \mathbb{R}^m)\\
G\colon [0,\Tstar] \times \mathbb{R}^+ \times (\mathbb{R}\times H^1) \to \mathbb{R}\\
L\colon [0,\Tstar] \times \mathbb{R}^+ \times (\mathbb{R}\times H^1) \to \mathbb{R}\\
\end{array}
\end{equation}
as
\begin{equation}
\label{eq:explicitFBGL}
\begin{array}{rl}
F(t,K,S,X)= & \left . I(X) \Bigg (\frac{1}{2} {X} |u_t(K,S,X)|^2 + \right .\\
&+\frac{1}{2} \left\langle u_t(K,S,X), \partial_x u_t(K,S,X) \right\rangle I(X) +\\ 
&+ 2 \left\langle u_t(K,S,X), \partial_x u_t(K,S,X) \right\rangle - \theta_t(K,S,X) \partial_x u_t^{(1)}(K,S,X)\ell \Bigg ) +\\
&+ { X} \left ( |u_t(K,S,X)|^2 - \theta_t(K,S,X) u_t^{(1)}(K,S,X) \right ) - \\ 
&\left . - 2 \left\langle \theta_t(K,S,X)\ell + u_t(K,S,X) \ln\left( \frac{K}{S} \right ), (\partial_x u_t(K,S,X)) \ln\left( \frac{K}{S} \right ) \right\rangle \right .\\
B(t,K,S,X)= & 2 { X} u_t^*(K,S,X) + 2 (\partial_x u_t^*(K,S,X)) I(X) \\
L(t,K,S,X)= & X[0] - \sum_{j=2}^m \left ( u_t^{(j)}(K,S,X)[0]\right )^2 \ln^2\left ( \frac{K}{S} \right )\\
G(t,K,S,X)= & u_t^{(1)}(K,S,X)[0] \ln\left ( \frac{K}{S} \right )
\end{array}
\end{equation}
where $\theta_t$ in the expression for $F$ is defined in (\ref{eq:theta}).
Under Hypothesis \ref{hp:suu} $F$, $B$, $G$ and $L$ are locally Lipschitz in $S,X$ uniformly in $t$, moreover, for all $K>0$ there exists a $M>0$ such that
\begin{equation}
\label{xlineargrowth}
\left \{
\begin{array}{l}
|F(t,K,S,X)|_{H^1} + |B(t,K,S,X)|_{(H^1)^m} \leq M(1+ |X|_{H^1})\\
G(t,K,S,X)\leq M
\end{array}
\right .
\end{equation}
for all $t$.

\noindent
\textbf{We fix now a $K>0$, and  avoid to write, from now, the dependence on $K$.}

\noindent
Using such a notation the (\ref{eq:stataoriginale}) that can be rewritten as:
\begin{equation}
\left \{
\label{eqstateinfdim}
\tag{EQ}
\begin{array}{rl}
\ud X_t = & AX_t + F(t,S_t,X_t) \ud t + B(t,S_t,X_t) \ud W_t , \;\;\; X_0=x_0>0 \\
\ud S_t= & \left ( \sqrt{L(t,S_t,X_t)} - G(t,S_t,X_t) \right )  S_t \ud W_t^{(1)}, \;\;\; S_0=s_0>0
\end{array}
\right .
\end{equation}
We call $\bar F$ the function defined changing in the definition of $F$ $\theta_t(K,S,X)$ with $\bar\theta_t(K,S,X)$. Note that $\bar F$ is locally Lipschitz in $S,X$ uniformly in $t$ and satisfies (\ref{xlineargrowth}).

\begin{Notation}
We will use the notation $e^{tA}$ instead of $T(t)$ defined in (\ref{eq:defsemigroup}).
\end{Notation}

\noindent
From the general theory (see \cite{DaPratoZabczyk92}) we have:
\begin{Definition}
\label{defsoluzione}
An $H^1\times \mathbb{R}$- valued predictable process $(X_t,S_t)$, $t\in [0,\Tstar]$ is called (mild) solution of (\ref{eqstateinfdim}) if 
\begin{equation}
\label{eq:conditionforsolution}
\mathbb{P} \left [ \int_0^{\Tstar} |(X_s,S_s)|_{H^1\times\mathbb{R}} \ud s <\infty \right ]=1
\end{equation}
and for an arbitrary $t\in [0,\Tstar]$ we have
\[
\left (
\begin{array}{l}
X_t\\
S_t
\end{array}
\right ) 
=
\left(
\begin{array}{l}
e^{tA} x_0 + \int_0^t e^{(t-s)A} \bar F(s,S_s,X_s) \ud s + \int_0^t e^{(t-s)A} B(s,S_s,X_s) \ud W_s\\
s_0+\int_0^t \left (  \sqrt{L(s,S_s,X_s)} - G(s,S_s,X_s) \right ) S_s \ud W_s^{(1)}
\end{array}
\right)
\]
\end{Definition}
\noindent
Note that this implies $L(s,S_s,X_s)\geq 0$.

\section{Results for (\ref{eqstateepsilon})}
\label{sec:Xepsilon}
We consider the approximating $X^\varepsilon_t(k)$ substituting 
$\sqrt{L(t,S_t,X_t)}$ in the second equation with $\sqrt{|L(t,S_t,X_t)|\vee \varepsilon}$:
\begin{equation}
\label{eqstateepsilon}
\tag{|EQ|$_\varepsilon$}
\left \{
\begin{array}{rl}
\ud X^{\varepsilon}_t = & AX^{\varepsilon}_t + {\bar F}(t,S^{\varepsilon}_t,X^{\varepsilon}_t) \ud t + B(t,S^{\varepsilon}_t,X^{\varepsilon}) \ud W_t, \;\;\;X^{\varepsilon}_0=x_0>0 \\
\ud S^{\varepsilon}_t= & \left ( \sqrt{|L(t,S_t,X^{\varepsilon}_t)|\vee \varepsilon} - G(t,S^{\varepsilon}_t,X^{\varepsilon}_t) \right ) S^{\varepsilon}_t \ud W_t^{(1)}, \;\;\; S_0=s_0
\end{array}
\right .
\end{equation}

The definition of solution of the (\ref{eqstateepsilon}) is analogous to the Definition \ref{defsoluzione}.

\begin{Notation}
We take a cut-off $\psi(\cdot)\colon \mathbb{R}\to \mathbb{R}$. In particular we assume that: $\psi(\cdot)$ is $C^{\infty}$, $\psi_{\big |[-1,1]}\equiv 1$ and that $\psi_{\big |(-\infty,-2) \cup (2,+\infty)}\equiv 0$.
\end{Notation}

\begin{Lemma}
\label{lemmastoppingtime}
Fix now $\bar\varepsilon>0$. Let $\varepsilon\in [0,\bar\varepsilon]$. Suppose that there exists a solution $(X^\varepsilon, S^\varepsilon)$ for (\ref{eqstateepsilon}) (that for $\varepsilon=0$ is (\ref{eqstateinfdim})). Then if we call $\tau_N$ the exit time defined as
\[
\tau_N=\inf \{ t\in [0,\Tstar] \; : \; |X^\varepsilon_t| >N \}\footnote{We use $|X_t|$ or $|X|$ for the norm in $H^1$ whereas $|X[0]|$ is the norm in $\mathbb{R}$}
\]
(and $+\infty$ if the set is void) we have that
\begin{equation}
\label{eq:lim}
\lim_{N\to +\infty} \mathbb{P} \left [ \tau_N \leq \Tstar \right ] = 0
\end{equation}
and the limit is uniformly in $\varepsilon \in [0,\bar\varepsilon]$ and in $(X^\varepsilon, S^\varepsilon)$.
\end{Lemma}
\begin{proof}
We call
\[
{\bar F}_N(t,S,X) := {\bar F}(t,S,X)\psi\left ( \frac{|X|}{N} \right )
\]
\[
B_N(t,S,X) := B(t,S,X)\psi\left ( \frac{|X|}{N} \right ).
\]
We choose $N>|X_0|$, we have
\[
X^\varepsilon_{t\wedge \tau_N} = e^{{t\wedge \tau_N}A} x_0 + \int_0^{t\wedge \tau_N} e^{(t-s)A}  {\bar F}_N(s,S_s,X^\varepsilon_{s\wedge \tau_N}) \ud s + \int_0^{t\wedge \tau_N} e^{(t-s)A} B_N(s,S_s,X^\varepsilon_{s\wedge \tau_N}) \ud W_s
\]
and so, using (\ref{xlineargrowth}) and Lemma 7.3 \cite{DaPratoZabczyk92},
\[
\mathbb{E}\left [ \sup_{s\in[0,t]} \left | X^\varepsilon_{t\wedge \tau_N} \right |^2 \right ] \leq C_{\Tstar} \left ( 1 + \mathbb{E}\int_0^t |X^\varepsilon_{s\wedge \tau_N}|^2 \ud s + \mathbb{E}\int_0^t |X^\varepsilon_{s\wedge \tau_N}|^2 \ud s  \right )
\]
where $C_{\Tstar}$ depends on $\Tstar$ and on the initial datum $x_0\in H^1$. So, thanks to Gronwall's lemma we have
\[
\mathbb{E}\left [ \sup_{s\in[0,t]} \left | X^\varepsilon_{t\wedge \tau_N} \right |^2 \right ] \leq C
\]
uniformly in $N$. In particular, since $X^{\varepsilon}_t$ is continuous (\cite{DaPratoZabczyk92} Theorem 7.4 \footnote{Once we have fixed $S^\varepsilon$, the solution of the (\ref{eqstateepsilon}) for $X^\varepsilon$ is unique and it satisfies the properties ensured by Theorem 7.4 of \cite{DaPratoZabczyk92}} then
\[
\sup_{0\leq t \leq \tau_N} |X^\varepsilon_t|^2 = N^2 \;\;\; on \; \{ \tau_N \leq \Tstar \}
\]
and then
\[
\mathbb{P}(\tau_N \leq \Tstar) \leq \frac{C}{N^2}
\]
and then we have the claim.
\qed
\end{proof}

\begin{Lemma}
\label{lm:existenceanduniquenessforepsilon}
The equation (\ref{eqstateepsilon}) has a unique solution (Definition \ref{defsoluzione}) $(X^\varepsilon_t, S^\varepsilon_t)$. Moreover $(X^\varepsilon_t, S^\varepsilon_t)$ belongs to $C([0,\Tstar];L^2(\Omega,\mathcal{F}, \mathbb{P};(H^1\times\mathbb{R}))$ and has continuous trajectories.
\end{Lemma}
\begin{proof}
We proceed localizing the problem using $\tau_N$ as defined in Lemma \ref{lemmastoppingtime} using the same notations $B_N$, ${\bar F}_N$ (and also $G_N(t,S,X) = G(t,S,X)\psi\left ( \frac{|X|}{N} \right )$, $L_N(t,S,X) = L(t,S,X)\psi\left ( \frac{|X|}{N} \right )$). The equation
\begin{equation}
\label{eq:epsilonN}
\left \{
\begin{array}{rl}
\ud X^{\varepsilon,N}_t = & AX^{\varepsilon,N}_t + {\bar F}_N(t,S^{\varepsilon,N}_t,X^{\varepsilon,N}_t) \ud t + B_N(t,S^{\varepsilon,N}_t,X^{\varepsilon,N}_t) \ud W_t \\
\ud S^{\varepsilon,N}_t= & \left (\sqrt{|L_N(t,S_t,X^{\varepsilon,N}_t)|\vee \varepsilon} - G_N(t,S^{\varepsilon,N}_t,X^{\varepsilon,N}_t) \right ) S^{\varepsilon,N}_t \ud W^{(1)}_t
\end{array}
\right .
\end{equation}
(with initial data $X^{\varepsilon,N,N}_0=x_0>0$ and $S_0=s_0>0$)
satisfies the hypotheses of Theorem 7.4 of \cite{DaPratoZabczyk92} and then has a unique continuous solution $(X^{\varepsilon,N}_t,S^{\varepsilon,N}_t)$ in $C([0,\Tstar], L^2(\Omega,\mathcal{F}, \mathbb{P};(H^1\times\mathbb{R})))$.

If $N'>N$ we have that $B_N=B_{N'}$ on $\{ |X|\leq N\}$ (and in the same way ${\bar F}_N={\bar F}_{N'}$, $G_N=G_{N'}$, $L_N=L_{N'}$ on $\{ |X|\leq N\}$) and then $X^{\varepsilon,N}_t=X^{\varepsilon,N'}_t$ on $\{ \tau_N >\Tstar \}$ a.s. So we can define 
\begin{equation}
\label{eq:defXepsilon}
X^\varepsilon_t=X^{\varepsilon,N}_t \qquad \text{on } \{ \tau_N >\Tstar \}\times [0,\Tstar].
\end{equation}
We can obtain an estimate as (\ref{eq:lim}) uniformly in $N$ and then ensure that $\lim_{N\to+\infty} \{ \tau_N >\Tstar \} =\Omega$. Note that, since $S^{\varepsilon}_{t\wedge \tau_N}$ solves
\[
S^{\varepsilon}_{t \wedge \tau_N} = s_0 + \int_0^{t \wedge \tau_N} (\sqrt{|L_N(t,S_t,X^{\varepsilon}_{t\wedge \tau_N})|\vee \varepsilon} - G_N(s,X^{\varepsilon}_{t\wedge \tau_N},S^{\varepsilon}_{t\wedge \tau_N}) S_s \ud W^{(1)}_s
\]
then we have
\begin{equation}
\label{estimateonS}
\mathbb{E} \left [ |S^{\varepsilon}_{t\wedge \tau_N}|^2  \right ] \leq s_0 e^{2 (N+N^2) t}
\end{equation}
where the second term does not depend on $\varepsilon$ and then $\mathbb{P} \left [ \int_0^{\Tstar} |S^{\varepsilon}_{t}|^2  \ud t <+\infty \right ] =0$ as required by (\ref{eq:conditionforsolution}). The uniqueness follows from the uniqueness for the localized problems. The regularity properties follow from the regularity for the approximating equations.
\qed
\end{proof}

\begin{Lemma}
\label{lm:integralformforepsilon}
Consider $(X^\varepsilon_t,S^{\varepsilon}_t)$ as in Lemma \ref{lm:existenceanduniquenessforepsilon}, then $X^\varepsilon_t$ it is a solution of the following integral equation in $C([0,\Tstar]; L^2(\Omega,\mathcal{F},\mathbb{P};L^2(0,T^*)))$:
\begin{equation}
\label{eq:integral-eps}
X^\varepsilon_t = x_0 + \int_0^t \partial_x X^\varepsilon_s[x] \ud s + \int_0^t {\bar F}(s,S^{\varepsilon}_t,X^\varepsilon_s) \ud s + \int_0^t B(s,S^{\varepsilon}_t,X^\varepsilon_s) \ud W_s
\end{equation}
\end{Lemma}
\begin{proof}
We can assume that ${\bar F}$ and $B$ are Lipschitz-continuous in $S^\varepsilon\in\mathbb{R}$ and $X^\varepsilon\in H^1$ uniformly in $t$ and $\omega$ (with Lipschitz constant $C$), otherwise we can localize the problem as in the proof of Lemma \ref{lm:existenceanduniquenessforepsilon}.

Consider the Yosida approximation of $A$ given by $A_n= n^2 (nI - A)^{-1} - nI$. We consider the solution $X^{\varepsilon,n}_t$ of the equation
\[
\ud X^{\varepsilon,n}_t = A_n X^{\varepsilon,n}_t \ud t + {\bar F}(t,S^{\varepsilon}_t, X^{\varepsilon,n}_t) \ud t + B(t, S^{\varepsilon}_t, X^{\varepsilon,n}_t) \ud W_t, \;\;X^\varepsilon_0=x_0.
\]
Since $A_n$ is linear continuous, and then Lipschitz, the mild form of such an equation can be written in two equivalent ways:
\[
X^{\varepsilon,n}_t = e^{A_n t} x_0 + \int_0^t e^{(t-s)A_n} {\bar F}(s,S^{\varepsilon}_t, X^{\varepsilon,n}_t) \ud s + \int_0^t e^{(t-s)A_n} B(s, S^{\varepsilon}_t, X^{\varepsilon,n}_t) \ud W_s
\]
and
\begin{equation}
\label{eq:secondmild}
X^{\varepsilon,n}_t = x_0+ \int_0^t A_n X^{\varepsilon,n}_t \ud s + \int_0^t {\bar F}(s,S^{\varepsilon}_t, X^{\varepsilon,n}_t) \ud s + \int_0^t B(s, S^{\varepsilon}_t, X^{\varepsilon,n}_t) \ud W_s.
\end{equation}
Moreover (see \cite{DaPratoZabczyk92} Proposition 7.5 page 193)
\begin{equation}
\label{eq:convXepsilonnXepsilon}
X^{\varepsilon,n}\xrightarrow[C(\lbrack0,\Tstar\rbrack; L^2(\Omega,\mathcal{F},\mathbb{P};H^1(0,T^*)))]{n\to\infty} X^\varepsilon.
\end{equation}
In order to prove the claim we need only to check that every term of (\ref{eq:secondmild}) converges to the corresponding term of the (\ref{eq:integral-eps}) in $C([0,\Tstar]; L^2(\Omega,\mathcal{F},\mathbb{P};L^2(0,T^*)))$:
\begin{multline}
\sup_{t\in [0,\Tstar]} \mathbb{E} \left [ \left | \int_0^t B(s,S^{\varepsilon}_t, X^\varepsilon_s) - B(s, S^{\varepsilon}_t, X^{\varepsilon,n}_t) \ud W_s \right |_{L^2}^2 \right ] \leq \\
\sup_{t\in [0,\Tstar]} \mathbb{E} \left [ \left | \int_0^t B(s,S^{\varepsilon}_t, X^\varepsilon_s) - B(s, S^{\varepsilon}_t, X^{\varepsilon,n}_t) \ud W_s \right |_{H^1}^2 \right ] \leq \\
\leq C_1 \mathbb{E} \left [ \int_0^{\Tstar} \left | B(s,S^{\varepsilon}_t, X^\varepsilon_s) - B(s, S^{\varepsilon}_t, X^{\varepsilon,n}_t) \right |_{H^1}^2 \ud s \right ] \leq \\
\leq C_1 C^2 \mathbb{E} \left [ \int_0^{\Tstar}  \left |X^\varepsilon_s- X^{\varepsilon,n}_t \right |_{H^1}^2 \ud s \right ] \xrightarrow{n\to\infty} 0 
\end{multline}
where the last convergence holds since we have (\ref{eq:convXepsilonnXepsilon}). The estimation with the term with ${\bar F}$ can be done in the same way. Moreover
\begin{multline}
\sup_{t\in [0,\Tstar]} \mathbb{E} \left [ \int_0^{t} \left | \partial_x X^\varepsilon_s[x] - A_n X^{\varepsilon,n}_t[x]  \right |_{L^2}^2 \ud s \right ] \leq \\
\leq I_1 + I_2  \nd \sup_{t\in [0,\Tstar]} \mathbb{E} \left [ \int_0^{t} \left | \partial_x A_n X^\varepsilon_s[x] - A_n X^{\varepsilon,n}_t[x]  \right |_{L^2}^2 \ud s \right ] +\\
+\sup_{t\in [0,\Tstar]} \mathbb{E} \left [ \int_0^{t} \left | \partial_x X^\varepsilon_s[x] - \partial_x X^{\varepsilon,n}_t[x]  \right |_{L^2}^2 \ud s \right ].
\end{multline}
For $I_2$ we have:
\[
I_2\leq \mathbb{E} \left [ \int_0^{\Tstar} \left | \partial_x X^\varepsilon_s[x] - \partial_x X^{\varepsilon,n}_t[x] \right |_{L^2}^2 \ud s \right ] \leq \mathbb{E} \left [ \int_0^{\Tstar}  \left |X^\varepsilon_s- X^{\varepsilon,n}_t \right |_{H^1}^2 \ud s \right ] \xrightarrow{n\to\infty}0,
\]
where we used that $\left | \partial_x X^\varepsilon_s[x] - \partial_x X^{\varepsilon,n}_t[x] \right |_{L^2} \leq \left |X^\varepsilon_s- X^{\varepsilon,n}_t \right |_{H^1}^2$ since the derivative is a linear continuous contractive function from $H^1$ to $L^2$.
To treat $I_1$ we have only to observe that $A_n \xrightarrow[n\to\infty]{\mathcal{L}(H^1;L^2)}$ and $X^{\varepsilon,n}\xrightarrow[n\to\infty]{C([0,\Tstar]; L^2(\Omega,\mathcal{F},\mathbb{P};H^1))} X^\varepsilon$. And so we have the claim.
\qed
\end{proof}

We consider now the stopping time 
\[
{\bar\tau}^\varepsilon := \inf\{ t\in [0,\Tstar] \; : \; L(t,S^{\varepsilon}_t,X_t^\varepsilon) < \varepsilon \}.
\]
and the process
\[
(\hat X^\varepsilon_t , \hat S^{\varepsilon}_t) := \chi_{[0,{\bar\tau}^\varepsilon]} (t) (X^\varepsilon_t ,  S^{\varepsilon}_t)
\]
It solves the integral equation
\begin{equation}
\label{eq:inteqsolvedbyXhatepsilon}
\left \{
\begin{array}{ll}
\hat X^\varepsilon_t = \chi_{[0,{\bar\tau}^\varepsilon]}(t) e^{tA} x_0 &+ \chi_{[0,{\bar\tau}^\varepsilon]}(t) \int_0^t \chi_{[0,{\bar\tau}^\varepsilon]}(s) e^{(t-s)A}  \bar F(s,\hat X^\varepsilon_t , \hat S^{\varepsilon}_t) \ud s+\\
&+ \chi_{[0,{\bar\tau}^\varepsilon]}(t) \int_0^t \chi_{[0,{\bar\tau}^\varepsilon]}(s) e^{(t-s)A}   B(s,\hat X^\varepsilon_t , \hat S^{\varepsilon}_t) \ud W_s\\
\multicolumn{2}{l}{S^{\varepsilon}_t = \chi_{[0,{\bar\tau}^\varepsilon]}(t) s_0+ \chi_{[0,{\bar\tau}^\varepsilon]}(t) \int_0^t \chi_{[0,{\bar\tau}^\varepsilon]}(s) \left (  \sqrt{L(s,\hat X^\varepsilon_t , \hat S^{\varepsilon}_t)} - G(s,\hat X^\varepsilon_t , \hat S^{\varepsilon}_t) \right ) \hat S^{\varepsilon}_t \ud W_s^{(1)}}
\end{array}
\right .
\end{equation}
Given a $\gamma<\varepsilon$ we have that
\[
(\hat X_t^\varepsilon,  \hat S_t^\varepsilon) = (\hat X_t^\gamma, \hat S_t^\gamma) \qquad \text{on} \; t\leq {\bar\tau}^\varepsilon
\]
So, we can define 
\begin{equation}
\label{eq:defSX}
(  X_t,   S_t) := \lim_{\varepsilon \to 0} (  \hat X_t^\varepsilon, \hat   S_t^\varepsilon)
\end{equation}
on $t\leq {\bar\tau}$ where ${\bar\tau}$ is defined as
\[
{\bar\tau} := \sup_{\varepsilon >0},
\]
and $(  X_t,   S_t) := (0,0)$ on $t > {\bar\tau}$. Passing to the limit in (\ref{eq:inteqsolvedbyXhatepsilon}) we have that $(  X_t,   S_t)$ is a solution of 
\begin{equation}
\label{eq:inteqsolvedbyXhat}
\left \{
\begin{array}{ll}
 X_t = \chi_{[0,{\bar\tau}]}(t) e^{tA} x_0 &+ \chi_{[0,{\bar\tau}]}(t) \int_0^t \chi_{[0,{\bar\tau}]}(s) e^{(t-s)A}  \bar F(s, X_t ,  S_t) \ud s+\\
&+ \chi_{[0,{\bar\tau}]}(t) \int_0^t \chi_{[0,{\bar\tau}]}(s) e^{(t-s)A}   B(s, X_t ,  S_t) \ud W_s\\
\multicolumn{2}{l}{S_t = \chi_{[0,{\bar\tau}]}(t) s_0+ \chi_{[0,{\bar\tau}]}(t) \int_0^t \chi_{[0,{\bar\tau}]}(s) \left (  \sqrt{L(s, X_t ,  S_t)} - G(s, X_t ,  S_t) \right )  S_t \ud W_s^{(1)}}
\end{array}
\right .
\end{equation}
Moreover if we set 
\[
{\bar\tau} := \sup_{\varepsilon >0} {\bar\tau}^{\varepsilon} = \inf\{ t\in [0,\Tstar] \; : \; X_t[0] \leq 0 \},
\]
since $(  \hat X_t^\varepsilon, \hat   S_t^\varepsilon)$ is a solution of (\ref{eqstateinfdim}) until time ${\bar\tau}_\varepsilon$ and   (\ref{eqstateinfdim}) is locally Lipschitz until time ${\bar\tau}$, $(  X_t ,   S_t)$ is the only solution of (\ref{eqstateinfdim}) until ${\bar\tau}$.

From Lemma \ref{lm:integralformforepsilon} we obtain the following corollary:

\begin{Lemma}
\label{lm:integralformuntiltime}
$X_t$ defined in (\ref{eq:defSX}) is a solution of the following integral equation:
\begin{multline}
\label{eq:integral}
  X_t = \chi{[0,{\bar\tau}]}(t) \bigg ( x_0 + \int_0^t \partial_x \chi_{[0,{\bar\tau}]}(s)   X_s[x] \ud s + \int_0^t \chi_{[0,{\bar\tau}]}(s) {\bar F}(s,  S_s,  X_s) \ud s +\\
+ \int_0^t \chi_{[0,{\bar\tau}]}(s) B(s,  S_s,  X_s) \ud W_s \bigg )
\end{multline}
\end{Lemma}
We call $Y_t$ the process
\begin{multline}
\label{eq:defYt}
Y_t = x_0 + \int_0^t \chi_{[0,{\bar\tau}]}(s) \partial_x   X_s[x] \ud s + \int_0^t \chi_{[0,{\bar\tau}]}(s) {\bar F}(s,  S_s,  X_s) \ud s +\\
+ \int_0^t \chi_{[0,{\bar\tau}]}(s) B(s,  S_s,  X_s) \ud W_s
\end{multline}

\section{The process $\xi_t$ and its properties}
\label{sec:xi}
In this section we (re)introduce the process $\xi_t$ that we used in the introduction as starting point. We will prove that such a process, here defined integrating $X_t$, indeed solves the SDE (\ref{eq:BLforulation}) that appears in \cite{BraceGoldys01}.

Consider now the real process $\left ( \xi_t^T \right )_{t\in[0,T^*]}$ where
\begin{equation}
\label{eq:defxi}
\xi_t^T = \int_0^{T-t} Y_t[x] \ud x.
\end{equation}
If we define 
\[
\begin{array}{l}
\Phi^T\colon [0,\Tstar]\times L^2(0,T^*) \to \mathbb{R}\\
\Phi^T \colon (t,\phi)\mapsto \int_0^{T-t} \phi[x] \ud x.
\end{array}
\]
we have that $\xi_t^T = \Phi^T(t,X_t)$. We want to find a (real) SDE solved by $\xi_t^T$ and so we apply Ito's formula to $\Phi^T$ using the fact that $X_t$ satisfies (\ref{eq:integral}). We have to put some attention because $\partial_t\Phi^T$ is not defined on all $[0,\Tstar]\times L^2(0,T^*)$:

\begin{Proposition}
\label{prop:xisolvestheequation-prima}
$\xi_t^T$ solves the following SDE:
\begin{equation}
\label{eq:sdeofxi-prima}
\left \{
\begin{array}{rcl}
\xi_{t\wedge{\bar\tau}}^{T} & = & \int_0^{t} \chi_{[0,{\bar\tau}]}(s) \left ( \xi_s^{T} \left ( \left ( 1+ \frac{1}{4} \xi_s^{T} \right ) |v_s|^2 -  \theta_s v_s^{(1)}  \right ) \right ) \ud s - \\
& & - \int_0^{t} \chi_{[0,{\bar\tau}]}(s) \left (  \theta_s + v_s^{(1)} \ln \left( \frac{k}{  S_s} \right) \right )^2 \ud s -\\ 
& & - \int_0^{t} \chi_{[0,{\bar\tau}]}(s) \sum_{j=2}^{m} \left ( v_s^{(j)} \right )^2 \ln^2 \left( \frac{k}{  S_s} \right ) \ud s + \int_0^{t} \chi_{[0,{\bar\tau}]}(s) 2 \xi_s^{T} v_s^* \ud W_s\\
\multicolumn{3}{l}{\xi_0^{T}=\int_0^{T} Y_0[x] \ud x =\int_0^{T} x_0[x] \ud x}
\end{array}
\right .
\end{equation}
where we called $v_s=u_s(k,  S_s,   X_s)[T-s]$ and $ \theta_t$ is defined in (\ref{eq:theta}) (note that on $t\leq{\bar\tau}$ we have that $\theta_t=\bar\theta_t$).
\end{Proposition}
\begin{proof}
We want to apply Ito's formula to $\Phi^T$ along $Y_t$ using the fact that $Y_t$ satisfies integral equation (\ref{eq:defYt}) and then it is in $C([0,\Tstar]; L^2(\Omega,\mathcal{F},\mathbb{P};L^2(0,T^*)))$. We have that
\[
\partial_\phi \Phi^T(t,\phi) = \chi_{[0,T-t]}
\]
and 
\[
\partial^2_\phi \Phi^T(t,\phi) = 0.
\]
and then they are uniformly continuous on bounded set. More problematic is the derivative along the time variable. Anyway we can observe that for $t\leq{\bar\tau}$ we have that $Y_t =   X_t$, and for $t\leq {\bar\tau}$ $  X_t$ is pathwise continuous in $H^1$. So we have that, for $t\leq {\bar\tau}$
\[
\partial_t \Phi^T(t,Y_t) = -  X_t[T-t].
\]
$\partial_t \Phi^T$ is not defined on all $[0,\Tstar]\times L^2(0,T^*)$ (and a fortiori it is not uniformly continuous on bounded subsets of $[0,\Tstar]\times L^2(0,T^*)$), we can anyway observe that what is really needed in the proof of Ito's formula (see \cite{DaPratoZabczyk92} Theorem 4.17 page 105) is just the convergence
\begin{equation}
\label{eq:Itoconv}
\sum_{j=0}^{M-1} \Phi^T(t_{j+1}, Y_{t_{j+1}}) - \Phi^T(t_{j}, Y_{t_{j+1}}) \xrightarrow{\flat(\pi)\to 0} \int_0^t \partial_t \Phi^T (s, Y_s) \ud s \;\;\; \mathbb{P}-a.s.
\end{equation}
where ``$\flat(\pi)\to 0$'' means that the mesh $\flat(\pi)$ of the partition $\pi=\{ 0=t_0<t_1<...<t_{M-1}<t_M=t \}$ goes to zero
Note that we only need to verify it on $t\leq {\bar\tau}$, thanks to the form of equation (\ref{eq:sdeofxi-prima}) and to the fact that the derivative w.r.t. the time enters only in the deterministic integral so (\ref{eq:Itoconv}) can be proved in our case since, for $t\leq {\bar\tau}$,
\[
\Phi^T(t_{j+1}, Y_{t_{j+1}}) - \Phi^T(t_{j}, Y_{t_{j+1}}) = 
\int_{T-t_j}^{T-t_{j+1}} -   X_{t_{j+1}}[x] \ud x
\]
and, the trajectories of $  X_t(\omega)$ for $t\leq{\bar\tau}$ are continuous in $H^1$.

So we can now use Ito formula end we find:
\begin{multline}
\xi_{t\wedge{\bar\tau}}^T = \Phi^T(t,Y_t) = \Phi^T(0,x_0) + \int_0^{t} \chi_{[0,{\bar\tau}]}(s) \left ( \partial_t \Phi^T(s,Y_s) + \left\langle \partial_x \Phi^T(s,Y_s), \partial_x   X_s[\cdot]  \right\rangle_{L^2} \right ) \ud s +\\
+ \int_0^{t} \chi_{[0,{\bar\tau}]}(s)  \left\langle \partial_x \Phi^T(s,Y_s), {\bar F}(s,{  S}_s,  X_s)  \right\rangle_{L^2} \ud s + \\
+ \int_0^{t} \chi_{[0,{\bar\tau}]}(s) \frac{1}{2} Tr [B(s,{  S}_s,   X_s) \partial^2_x \Phi^T(s,Y_s) B(s,{  S}_s,   X_s)] \ud s + \\
+\int_0^{t} \chi_{[0,{\bar\tau}]}(s) \left\langle \partial_x \Phi^T(s,Y_s), B(s,{  S}_s,   X_s)   \ud W_s \right\rangle_{L^2}= \\
= 0 + \int_0^{t\wedge{\bar\tau}} \left ( -   X_s[T-s] + \int_0^{T-s} \partial_x    X_s[x] \ud x \right ) \ud s + \\
+ \int_0^{t} \chi_{[0,{\bar\tau}]}(s) \int_0^{T-s} {\bar F}(s,{  S}_s,  X_s)[x] \ud x \ud s + \int_0^t \chi_{[0,{\bar\tau}]}(s) \int_0^{T-s} B(s,{  S}_s,   X_s)[x] \ud x \ud W_s =
\end{multline}
using the explicit expression of ${\bar F}$ and $B$ given by (\ref{eq:explicitFBGL})
\begin{multline} 
= \int_0^{t\wedge{\bar\tau}} -{  X}_s[0] \ud s +\\
+ \int_0^{t\wedge{\bar\tau}} \left | I({  X}_s)[x] \left ( \left (1+ \frac{1}{4} I({  X}_s)[x] \right ) |u_s(k,{  S}_s,{  X}_s)[x]|^2 - \theta_s^{(1)} u^{(1)}_s(k,{  S}_s, {  X}_s)[x] \right ) \right |_{x=0}^{x=T-s} \ud s- \\
- \int_0^{t\wedge{\bar\tau}} \left | \left |\theta_s\ell + u_s(k,{  S}_s,{  X}_s)[x] \ln\left( \frac{k}{{  S}_s} \right)\right |^2 \right |_{x=0}^{x=T-s} \ud s+ \\
+ \int_0^t \chi_{[0,{\bar\tau}]}(s) \left | 2 I({  X}_s)[x] u^*_s(k,{  S}_s,{  X}_s)[x] \right |_{x=0}^{x=T-s} \ud W_s=
\end{multline}
noting that, from (\ref{eq:theta}), $\left |\theta_s\ell + u_s(k,S_s,X_s)[x] \ln\left( \frac{k}{S_s} \right)\right |^2_{x=0} =Y_s[0]$, that for $s\leq {\bar\tau}$ is equal to $  X_s[0]$, and recalling that $Y_s=  X_s$ on $s\leq {\bar\tau}$ we have
\begin{multline}
= \int_0^t \chi_{[0,{\bar\tau}]}(s) \xi_s^T \left ( \left (1+ \frac{1}{4} \xi_s^T \right ) |v_s|^2 - \theta_s^{(1)} v_s \right )  \ud s - \int_0^t \chi_{[0,{\bar\tau}]}(s) \left |\theta_s \ell + v_s \ln\left( \frac{k}{S_s} \right)\right |^2 \ud s +\\
+ \int_0^t \chi_{[0,{\bar\tau}]}(s) 2 \xi_s^T v^*_s  \ud W_s
\end{multline}
That is the claim. 
\qed
\end{proof}
Since we obtained at the beginning $X_t[x]$ differentiating (formally) with respect to $x$ the process $t\mapsto \xi_t^{t+x}$ so what we have obtained now it not unexpected but exactly equation (\ref{eq:BLforulation}).

\begin{Theorem}
\label{th:existence-uniqueness-X}
For all $T\in [0,T^*]$, $\xi_t^T\geq 0$ for all $t\leq{\bar\tau}$.
\[
\mathbb{P} \{  {\bar\tau}\leq T^* \} =0.
\]
and $(S_t,X_t)=(S_t,Y_t)$, $0\leq t\leq T^*$ is the only solution of (\ref{eqstateinfdim}). Moreover it is in $C([0,\Tstar];L^2(\Omega,\mathcal{F}, \mathbb{P};(H^1\times\mathbb{R}))$
and has continuous trajectories.
\end{Theorem}
\begin{proof}
Since $\xi_{t\wedge{\bar\tau}}^T$ solves the (\ref{eq:sdeofxi-prima}) (note that we can use on the right side $\xi_{s\wedge{\bar\tau}}^T$ instead of $\xi_{s}^T$ without any difference) we can write it as
\[
\xi_{t\wedge{\bar\tau}}^T = e^{L_t^T} f(T,k) - e^{L_t^L} \int_0^{t\wedge{\bar\tau}} e^{-L_s^T} \left |  \theta_s \ell + v_s\ln \left ( \frac{k}{  S_s} \right ) \right |^2 \ud s
\]
where
\[
L_t^T=L_t(T,K)= \int_0^t \chi_{[0,{\bar\tau}]}(s) 2 v_s^* \ud W_s - \int_0^t \chi_{[0,{\bar\tau}]}(s) \left ( \left ( 1- \frac{1}{4} \xi_{s\wedge{\bar\tau}}^T \right ) |v_s|^2 + \theta_s v_s^{(1)} \right ) \ud s.
\]
Since $\xi_{\bar\tau}^{\bar\tau}=0$ we have
\begin{multline}
f({\bar\tau},k)= \int_0^{\bar\tau} e^{-L_s^{\bar\tau}} \left | \theta_s\ell + v_s \ln \left (\frac{k}{  S_s} \right ) \right |^2  \ud s = f({\bar\tau},k)=\\
= \int_0^{t\wedge{\bar\tau}} e^{-L_s^{\bar\tau}} \left | \theta_s\ell + v_s \ln \left (\frac{k}{  S_s} \right ) \right |^2  \ud s + \int_{t\wedge{\bar\tau}}^{\bar\tau} e^{-L_s^{\bar\tau}} \left | \theta_s\ell + v_s \ln \left (\frac{k}{  S_s} \right ) \right |^2  \ud s
\end{multline}
Hence
\begin{equation}
\label{eq:xipositive}
\xi_{t\wedge{\bar\tau}}^{\bar\tau}= e^{L_t^{\bar\tau}} \int_{t\wedge{\bar\tau}}^{\bar\tau} e^{-L_s^{\bar\tau}} \left | \theta_s\ell + v_s \ln \left (\frac{k}{  S_s} \right ) \right |^2 \ud s \geq 0.
\end{equation}
In particular
\begin{equation}
\label{eq:xitauzero}
\int_0^{\bar\tau} x_0 [x] \ud x = \xi_{0}^{\bar\tau}= \int_{0}^{\bar\tau} e^{-L^{\bar\tau}_s} \left | \theta_s\ell + v_s \ln \left (\frac{k}{  S_s} \right ) \right |^2 \ud s = \int_{0}^{\bar\tau} e^{-L^{\bar\tau}_s}   X_s[0] \ud s = 
\end{equation}
We take now $h\in (0,1)$, as before we can obtain
\begin{equation}
\label{eq:xihtauzero}
\int_0^{(1-h){\bar\tau}} x_0 [x] \ud x = \xi_{0}^{(1-h){\bar\tau}}=\int_{0}^{(1-h){\bar\tau}} e^{-L^{(1-h){\bar\tau}}_s}   X_s[0] \ud s
\end{equation}
From (\ref{eq:xitauzero}) and (\ref{eq:xihtauzero}) we obtain
\[
\frac{\int_{(1-h){\bar\tau}}^{\bar\tau} x_0 [x] \ud x}{h} = \frac{\int_{(1-h){\bar\tau}}^{\bar\tau} e^{-L^{\bar\tau}_s}   X_s[0] \ud s}{h} + \frac{\int_{0}^{(1-h){\bar\tau}} \left ( e^{-L^{\bar\tau}_s} - e^{-L^{(1-h){\bar\tau}}_s} \right )   X_s[0] \ud s
}{h}.
\]
Now assume by contradiction that there exist a subset $\bar\Omega\subseteq \Omega$ with $\mathbb{P}(\bar\Omega)=c>0$ such that ${\bar\tau}(\omega)\leq T^*$ for $\omega\in\bar\Omega$. Observe also that, since $x_0>0$ and it is continuous (it is in $H^1$), we have $x_0\geq c_2>0$. From the previous equation we obtain:
\begin{multline}
\frac{\int_{\bar\Omega} \left |\int_{(1-h){\bar\tau}}^{\bar\tau} x_0 [x] \ud x \right | \ud \mathbb{P}(\omega)}{h} \leq \frac{ \int_{\bar\Omega} \left | \int_{(1-h){\bar\tau}}^{\bar\tau} e^{-L^{\bar\tau}_s}   X_s[0] \ud s\right | \ud \mathbb{P}(\omega)}{h} + \\
+ \frac{\int_{\bar\Omega} \left |\int_{0}^{(1-h){\bar\tau}} \left ( e^{-L^{\bar\tau}_s} - e^{-L^{(1-h){\bar\tau}}_s} \right )   X_s[0] \ud s
\right |\ud \mathbb{P}(\omega) }{h}.
\end{multline}
Passing to the liminf in $h\to 0$ the left side is greater than $c_2 c \int_{\bar\Omega} |{\bar\tau}| \ud \mathbb{P}(\omega) >0$ while the right side goes to zero.

So we can see that $\chi_{[0,{\bar\tau}]}(t)$ in equation (\ref{eq:inteqsolvedbyXhat}) is indeed always $1$ for $t\in[0,T^*]$ and the process $X_t$ solves the integral equation 
\[
\left \{
\begin{array}{l}
 X_t =  e^{tA} x_0 + \int_0^t e^{(t-s)A}  \bar F(s, X_t ,  S_t) \ud s
+  \int_0^t e^{(t-s)A}   B(s, X_t ,  S_t) \ud W_s\\
S_t = s_0+  \int_0^t  \left (  \sqrt{L(s, X_t ,  S_t)} - G(s, X_t ,  S_t) \right )  S_t \ud W_s^{(1)}.
\end{array}
\right .
\]
Noting that for $t\leq T^*$ we have $t\leq\bar\tau$ and then $F=\bar F$ we have that $(X_t,S_t)$ is a solution of (\ref{eqstateinfdim}) on $[0,T^*]$. The regularity properties follow from the ones of $(X_t^\varepsilon,S_t^\varepsilon)$.
\qed
\end{proof}
\begin{Remark}
The argument used for the positivity of $\xi_t$ is the same used in \cite{BraceGoldys01}.
\end{Remark}

\begin{Proposition}
\label{prop:xisolvestheequation}
$\xi_t^T$ is positive and, on $[0,T^*]$, we have:
\begin{equation}
\label{eq:sdeofxi}
\left \{
\begin{array}{rcl}
\multicolumn{3}{l}{\ud S_t = S_t\theta_t \ud W_t^{(1)}}\\
\ud \xi_t^{T} & = & \xi_t^{T} \left ( \left ( 1+ \frac{1}{4} \xi_t^{T} \right ) |v_t|^2 - \theta_t v_t^{(1)}  \right ) \ud t - \left ( \theta_t + v_t^{(1)} \ln \left( \frac{k}{S_t} \right) \right )^2 -\\ 
& & - \sum_{j=2}^{m} \left ( v_t^{(j)} \right )^2 \ln^2 \left( \frac{k}{S_t} \right ) +2 \xi_t^{T} v_t^* \ud W_t\\
\multicolumn{3}{l}{\xi_0^{T}=\int_0^{T} X_0[x] \ud x}\\
\multicolumn{3}{l}{\left . \partial_T \xi_t^T \right |_{T=t} = \left ( \theta_t + v_t^{(1)} \ln \left( \frac{k}{S_t} \right) \right )^2 + \sum_{j=2}^{m} \left ( v_t^{(j)} \right )^2 \ln^2 \left( \frac{k}{S_t} \right ) }
\end{array}
\right .
\end{equation}
\end{Proposition}
\begin{proof}
This is just a corollary of Theorem \ref{th:existence-uniqueness-X}, the result follows from (\ref{eq:sdeofxi-prima}) using that $X_t=Y_t$ and that for $T^*\leq \bar\tau$. The feedback equation is easily seen to be satisfied thanks to the definition of $\theta_t$.
\qed
\end{proof}

\begin{Proposition}
\label{pr:positive}
$S_t$ is a strictly positive process, moreover, for all $x\in[0,T^*]$, we have that 
\[
X_t[x] \geq 0 \;\;\; \mathbb{P}-a.s.
\]
\end{Proposition}
\begin{proof}
The statement for $S_t$ is easy since it can be written as
\[
\frac{\ud S_t}{S_t} = \theta_t \ud W_t^{(1)}.
\]
The assertion for $X_t$ follows from the same arguments we used in Proposition \ref{prop:xisolvestheequation-prima} and in Theorem \ref{th:existence-uniqueness-X}. Indeed for $T_1< T_2\leq T^*$ we can define $\xi_t^{T_1,T_2}=\Phi^{T_1, T_2} (t,X_t)$ where
\[
\begin{array}{l}
\Phi^{T_1, T_2} \colon [0,T_2-T_1]\times L^2(0,T^*) \to \mathbb{R}\\
\Phi^{T_1, T_2} \colon (t,\phi)\mapsto \int_{T_1}^{T_2-t} X_t[x] \ud x.
\end{array}
\]
It can be seen using the same arguments we used in Proposition \ref{prop:xisolvestheequation-prima} that $\xi_t^{T_1,T_2}$ solves the equation
\begin{equation}
\label{eq:sdeofxi-T1}
\left \{
\begin{array}{rcl}
\ud \xi_t^{T_1,T_2} & = & \xi_t^{T_1,T_2} \left ( \left ( 1+ \frac{1}{4} \xi_t^{T_1,T_2} \right ) |v_t|^2 - \theta_t v_t^{(1)}  \right ) \ud t - \left ( \theta_t + v_t^{(1)} \ln \left( \frac{k}{S_t} \right) \right )^2 -\\ 
& & - \sum_{j=2}^{m} \left ( v_t^{(j)} \right )^2 \ln^2 \left( \frac{k}{S_t} \right ) +2 \xi_t^{T_1,T_2} v_t^* \ud W_t\\
\multicolumn{3}{l}{\xi_{0}^{T_1,T_2}=\int_{T_1}^{T_2} X_{0}[x] \ud x}
\end{array}
\right .
\end{equation}
Arguing as in Theorem \ref{th:existence-uniqueness-X} $\xi_t^{T_1,T_2}$ can be proved to be always positive in the interval $t\in [0,T_2-T_1]$ and then, step-by-step, to be always positive in the interval $t\in [0,T^*]$ and so we have that, for all $0\leq T_1\leq T_2 \leq T^*$ all the integrals $\int_{T_1}^{T_2} X_t[x] \ud x$ are positive and then, since $X_t\in H^1$, we have the claim.
\qed
\end{proof}

\section{Two examples}
\label{sec:examples}
We show now two possible examples in which the Hypotheses \ref{hp:suu}, \ref{hp:sqrtpositive} are satisfied. In both we take $m=2$ (similar examples with $m>2$ can be done). In both the cases $u^{(i)}$ depend on $\omega$ only through the variables $S$ and $X$.
Note that in both cases we choose a simple ``decoupled'' form for $u^{(i)}$ given by
\[
u_t^{(i)}(k,S,X)(\omega)[x] = \varphi^{(i)} \left ( X  \right ) \psi^{(i)} \left ( \frac{k}{S} \right ) \eta^{(i)}(|X[0]|).
\]

\subsection{First example: the volvol does not depend on $x$}
In this first example we assume $u_t(k,S,X)\colon x \mapsto u_t(k,S,X)[x]$ is a constant function. This means that $\partial_x u_t \equiv 0$ and we have not problems in satisfying points $(ii)$ $(iv)$ (and part of $(iii)$) of Hypothesis \ref{hp:suu}.

We take
\[
u_t^{(2)}(k,S,X)(\omega)[x] = \varphi^{(2)} \left ( \sup_{r\in[0,T^*]} \left | X[r] \right |  \right ) \frac{\psi^{(2)} \left ( \frac{k}{S} \right )}{\ln \left ( \frac{k}{S}\right )} \eta^{(2)}(X[0]) 
\]
for all  $x\in [0,T^*]$ and  $\omega\in\Omega$. We assume that
\[
\left \{
\begin{array}{l}
\eta^{(2)} \colon \mathbb{R}^+ \to \mathbb{R}\\
\eta^{(2)} \colon \sigma \mapsto \eta^{(2)}(\sigma)
\end{array}
\right .
\]
is locally Lipschitz, bounded, $\eta^{(2)}(0)=0$, and 
\begin{equation}
\label{eq:ex1-subsqrt}
\eta^{(2)}(\sigma) < \sqrt{\sigma} \qquad \text{ for all } \sigma >0.
\end{equation}
Moreover we assume that
\[
\left \{
\begin{array}{l}
\psi^{(2)} \colon \mathbb{R}^+ \to \mathbb{R}\\
\psi^{(2)} \colon \sigma \mapsto \psi^{(2)}(\sigma)
\end{array}
\right .
\]
is bounded and continuous and $\sigma \mapsto \frac{\psi^{(2)}(\sigma)}{\ln(\sigma)}$ is bounded and locally Lipschitz continuous. $\varphi^{(2)}$ is continuous, locally Lipschitz continuous and $\sigma\mapsto \varphi^{(2)}(\sigma)(1+\sigma)$ is bounded. Moreover
\begin{equation}
\label{eq:ex1-leq1}
\varphi^{(2)}(\sigma)\leq 1 \qquad\qquad \psi^{(2)}(\sigma) \leq 1
\end{equation}
We take
\[
u_t^{(1)}(k,S,X)(\omega)[x] = \varphi^{(1)} \left ( \sup_{r\in[0,T^*]} \left | X[r] \right |  \right ) \frac{\psi^{(1)} \left ( \frac{k}{S} \right )}{\ln \left ( \frac{k}{S}\right )} \eta^{(1)}(X[0]) 
\]
for all $x\in [0,T^*]$ and $\omega\in\Omega$. $\varphi^{(1)}$ is locally Lipschitz $\sigma\mapsto \varphi^{(2)}(\sigma)(1+\sigma)$ is bounded, $\psi^{(1)}$ is bounded and continuous and $\sigma \mapsto \frac{\psi^{(1)}(\sigma)}{\ln(\sigma)}$ is bounded and locally Lipschitz continuous, $\eta^{(1)}$ is locally Lipschitz continuous and bounded, $\sigma\mapsto \eta^{(1)}(\sigma) \sqrt{\sigma}$ locally Lipschitz continuous.

So if we define
\[
\tilde u_t^{(2)}(k,S,X) \nd u_t^{(2)} (k,S,X) \ln \left ( \frac{k}{S} \right ) = \varphi^{(2)} \left ( \sup_{r\in[0,T^*]} \left | X[r] \right |  \right ) \psi^{(2)} \left ( \frac{k}{S} \right ) \eta^{(2)}(X[0])
\]
we can note that (\ref{eq:ex1-subsqrt}) and (\ref{eq:ex1-leq1}) gives
\[
X[0] - \left ( \tilde u_t^{(2)}(k,S,X) \right )^2 \geq 0
\]
and it is equal to zero if and only if $X[0]=0$; so Hypothesis \ref{hp:sqrtpositive} is satisfied.

Part $(i)$ of Hypothesis \ref{hp:suu} follows by the boundedness of $\sigma \mapsto \psi^{(i)}(\sigma)$, of $\sigma \mapsto\varphi^{(i)}(\sigma)(1+\sigma)$ and of $\sigma \mapsto \eta^{(i)}$.

Local Lipschitz continuity properties required in $(iii)$ of Hypothesis \ref{hp:suu} follow by the local Lipschitz continuity properties of the functions considered.

\subsection{Second example: $u_t^{(i)}$ depends on $\xi$}

We assume now that $u_t(S,X)[x]$ depends on $X$ through $X[0]$ and $\int_0^x X[r] \ud r = \xi$, moreover $u_t(S,X)[x]$ depends on $x$ through $\int_0^x X[r] \ud r = \xi$, note that it is the quantity considered in the formulation of \cite{BraceGoldys01} and is the variable interesting from a financial point of view. Note that for technical reasons (to satisfy point $(iv)$ of Hypothesis \ref{hp:suu} we have to introduce a cut-off $\gamma_N$ in other example. $\gamma_N$ is a $C^{\infty}$ function $\mathbb{R}\to\mathbb{R}^+$ equal to $1$ in the interval $[-N,N]$ and equal to $0$ in $[2N, +\infty)$ and $(-\infty , -2N]$.

We assume 
\[
u_t^{(2)}(k,S,X)(\omega)[x] = \varphi^{(2)} \left ( \int_0^x X[r] \ud r \right ) \frac{\psi^{(2)} \left ( \frac{k}{S} \right )}{\ln \left ( \frac{k}{S}\right )} \eta^{(2)}(X[0]) \gamma_N(|X|_{H^1})
\]
for all  $x\in [0,T^*]$ and  $\omega\in\Omega$. We assume first that $\varphi^{(2)}\colon \mathbb{R} \to \mathbb{R}$ is $C^2$, bounded with first and second derivative bounded and we observe that the derivative $\frac{\ud}{\ud x} \left [ \varphi^{(2)} \left ( \int_0^x f[r] \ud r \right ) \right ]$ define a locally Lipschitz continuous function on $H^1$:
\begin{Lemma}
\label{lm:ex2-loclipschitz}
Suppose $\varphi^{(2)}\colon \mathbb{R}^+ \to \mathbb{R}$ is $C^2$, bounded with first and second derivative bounded: $|\varphi^{(2)}|\leq M$, $|\varphi^{(2)}_x|\leq M$, $|\varphi^{(2)}_{xx}|\leq M$. Then 
\begin{itemize}
\item[(a)] the function
\[
\left \{
\begin{array}{l}
\Gamma_{\varphi^{(2)}} \colon H^1(0,T^*; \mathbb{R}) \to H^1(0,T^*; \mathbb{R})\\
\Gamma_{\varphi^{(2)}} \colon f[\cdot] \mapsto \left ( x \mapsto \varphi^{(2)} \left ( \int_0^x f[r] \ud r \right ) \right )
\end{array}
\right .
\]
is locally Lipschitz continuous.
\item[(b)] the function
\[
\left \{
\begin{array}{l}
\Psi_{\varphi^{(2)}} \colon H^1(0,T^*; \mathbb{R}) \to H^1(0,T^*; \mathbb{R})\\
\Psi_{\varphi^{(2)}} \colon f[\cdot] \mapsto \left ( x \mapsto \varphi^{(2)}_x \left ( \int_0^x f[r] \ud r \right ) f[x] \right )
\end{array}
\right .
\]
is locally Lipschitz continuous.

\end{itemize}
\end{Lemma}
\begin{proof}
We prove only the point (b) because (a) is simpler and can be treated with the same arguments.

We take $f[\cdot]$ and $g[\cdot]$ in $H^1(0,T^*;\mathbb{R})$:
\begin{multline}
\left | \Psi_{\varphi^{(2)}} (f[\cdot]) - \Psi_{\varphi^{(2)}} (g[\cdot])
\right |_{H^1} = \\
=\left | \Psi_{\varphi^{(2)}} (f[\cdot]) - \Psi_{\varphi^{(2)}} (g[\cdot])
\right |_{L^2} + 
\left | \partial_x \left ( \Psi_{\varphi^{(2)}} (f[\cdot]) \right ) - \partial_x \left ( \Psi_{\varphi^{(2)}} (g[\cdot]) \right )
\right |_{L^2} =\\
= P_1 + P_2 \nd
\left | 
\left (\varphi^{(2)}_{x} \left ( \int_0^\cdot f[r] \ud r \right ) f[\cdot] - \varphi^{(2)}_x \left ( \int_0^\cdot g[r] \ud r \right ) g[\cdot] \right ) \right |_{L^2} + \\
+ \left | 
\varphi^{(2)}_{xx} \left ( \int_0^\cdot f[r] \ud r \right ) f^2[\cdot] + \varphi^{(2)}_{x} \left ( \int_0^\cdot f[r] \ud r \right ) f_x[\cdot] - \right .\\ 
\left . -\varphi^{(2)}_{xx} \left ( \int_0^\cdot g[r] \ud r \right ) g^2[\cdot] -\varphi^{(2)}_{x} \left ( \int_0^\cdot g[r] \ud r \right ) g_x[\cdot] \right |_{L^2}.
\end{multline}
We consider first $P_1$:
\begin{multline}
P_1 \leq P_1^1 + P_1^2 \nd \left | 
\varphi^{(2)}_{x} \left ( \int_0^\cdot f[r] \ud r \right ) \left ( f[\cdot] - g[\cdot] \right ) \right |_{L^2} + \\
+ \left | \left ( \varphi^{(2)}_x \left ( \int_0^\cdot f[r] \ud r \right ) - \varphi^{(2)}_x \left ( \int_0^\cdot g[r] \ud r \right ) \right ) g[\cdot] \right |_{L^2}.
\end{multline}
For $P_1^1$ we have simply $P_1^1 \leq M |f-g|_{L^2}\leq M |f-g|_{H^1}$. For $P_1^2$ note first that there exist a constant $C$ such that for every $h$ in a neighborhood (in $H^1$) of $g$ we have $|h|_{L^\infty}\leq C$, so locally we have
\begin{multline}
P_1^2 \leq MC \int_0^{\Tstar} \left ( \int_0^x f[r] - g[r] \ud r \right )^2 \ud x \leq MC \int_0^{\Tstar} \left ( \int_0^{\Tstar} f[r] - g[r] \ud r \right )^2 \ud x \leq \\
\leq \frac{CM}{{\Tstar}} \int_0^{\Tstar} \int_0^{\Tstar} (f[r] - g[r])^2 \ud r \ud x \leq CM |f-g|_{L^2} \leq CM |f-g|_{H^1}.
\end{multline}
We estimate now $P_2$:
\begin{multline}
P_2= P_2^1 + P_2^2 + P_2^3 + P_2^4 + P_2^5  \nd \left | \left ( 
\varphi^{(2)}_{xx} \left ( \int_0^\cdot f[r] \ud r \right ) -
\varphi^{(2)}_{xx} \left ( \int_0^\cdot g[r] \ud r \right ) \right ) f^2[\cdot] \right |_{L^2} +\\
+ \left | \left ( f[\cdot] - g[\cdot] \right ) f[\cdot] \varphi^{(2)}_{xx} \left ( \int_0^\cdot g[r] \ud r \right ) \right |_{L^2} +
\left | \left ( f[\cdot] - g[\cdot] \right ) g[\cdot] \varphi^{(2)}_{xx} \left ( \int_0^\cdot g[r] \ud r \right ) \right |_{L^2} +\\
+ \left | \left ( 
\varphi^{(2)}_{x} \left ( \int_0^\cdot f[r] \ud r \right ) - \varphi^{(2)}_{x} \left ( \int_0^\cdot g[r] \ud r \right ) \right ) f_x[\cdot] \right |_{L^2} +\left|\varphi^{(2)}_{x} \left ( \int_0^\cdot g[r] \ud r \right ) (f_x[\cdot]-g_x[\cdot]) \right |_{L^2}.
\end{multline}
Recalling that for $h$ in a neighborhood (in $H^1$) of $g$ and of $f$ $|h|_{L^\infty}\leq C$ we can estimate $P_2^2$ and $P_2^3$ as $P_1^1$, $P_2^1$ and $P_2^4$ can be treated as $P_1^2$. Eventually $P_2^5 \leq M |f_x-g_x|_{L^2}\leq M |f-g|_{H^1}$.
\qed
\end{proof}
We assume that: $\varphi^{(2)}$ is $C^2$ with bounded first and second derivative, $|\varphi^{(2)}|\leq 1$ and $\sigma\mapsto \varphi^{(2)}(\sigma)(1+|\sigma|)$ is bounded. Moreover we take $\psi^{(2)}$ bounded and continuous with $|\psi^{(2)}|\leq 1$ such that $\frac{\psi^{(2)}(\cdot)}{\ln(\cdot)}$ is bounded and locally Lipschitz continuous. $\eta^{(2)}$ is locally Lipschitz, with $\sigma\mapsto \eta^{(2)}(\sigma)(1+\sqrt{|\sigma|})$ bounded and locally Lipschitz continuous, $\eta^{(2)}(0)=0$, and 
\begin{equation}
\label{eq:ex2-subsqrt}
\eta^{(2)}(\sigma) < \sqrt{\sigma} \qquad \text{ for all } \sigma >0.
\end{equation}
We assume that $u_t^{(1)}$ has the following form:
\[
u_t^{(1)}(k,S,X)(\omega)[x] = \varphi^{(1)} \left ( \int_0^x X[r] \ud r \right ) \frac{\psi^{(1)} \left ( \frac{k}{S} \right )}{\ln \left ( \frac{k}{S}\right )} \eta^{(1)}(X[0]) \gamma_N(|X|_{H^1})
\]
where: $\varphi^{(1)}$ is a $C^2$ function with bounded first and second derivatives, $\sigma\mapsto \varphi^{(1)}(\sigma)(1+\sigma)$ is bounded; $\psi^{(1)}$ is bounded and continuous and $\sigma \mapsto \frac{\psi^{(1)}(\sigma)}{\ln(\sigma)}$ is bounded and locally Lipschitz continuous, $\eta^{(1)}$ is locally Lipschitz continuous and bounded with $\sigma\mapsto \varphi^{(1)}(\sigma)(1+\sqrt{|\sigma|})$ bounded and locally Lipschitz continuous.

We claim that the Hypotheses \ref{hp:suu} and \ref{hp:sqrtpositive} are satisfied. Hypothesis \ref{hp:sqrtpositive} follows by (\ref{eq:ex2-subsqrt}) and the fact that $\varphi^{(2)},\psi^{(2)} \leq 1$. Note that thank to the boundedness of $u_t^{(1)}$ and $u_t^{(2)}$ we have that $\theta \leq \sqrt{|X[0]|} +M$ for some constant $M$. So the first of $(iv)$ of Hypothesis \ref{hp:suu} follows by the boundedness of $x\mapsto \varphi^{(i)}(x)(1+\sqrt{|x|})$ and from the use of the cut-off $\gamma_N$; $(i)$ follows by the boundedness of $x\mapsto \varphi^{(i)}(x)(1+\sqrt{|x|})$, $x\mapsto \varphi^{(i)}(x)(1+|x|)$ and $x\mapsto \psi^{(i)}(x)$ and $(ii)$ from Lemma \ref{lm:ex2-loclipschitz} that gives also the local Lipschitz continuity property of $\partial_x u$ required in $(iii)$. The other local Lipschitz continuity property of $(iii)$ can be proved using the local Lipschitz continuity properties of $\psi^{(i)}$ , $\varphi^{(i)}$ and $\eta^{(i)}$ and by Lemma \ref{lm:ex2-loclipschitz}.

\section{Conclusions and future work}
\label{sec:varyingK}
We have proven an existence-and-uniqueness result for the implied volatility model presented in \cite{BraceGoldys01}.
The approach we used was based on the rewriting the problem in a suitable
Hilbert space formulation. We dealt with the one-parameter family of European call option $O_t(K,T)$ for a fixed strike price $K>0$. 

A natural object for future work is studying the case of the whole family $O_t(K,T)$ varying both the strike and the expiration time. As we have already observed in the introduction the feedback condition imply the unpleasant equation (\ref{eq:theta}) in which $\theta_t$ appears as a function of $K$. $\theta_t$ is the process that drives the evolution of the stock price and of course, if we want to deal with the general case of the complete family $O_t(K,T)$ varying both $T$ and $K$, it has be the same for every $K$. So we obtain the following family of \textit{compatibility conditions}: for all the strikes $K_1>0$ and $K_2>0$
\[
\theta_t(K_1,S_t,X_t(K_1))=\theta_t(K_2,S_t,X_t(K_2))
\]
that is
\begin{multline}
\label{eq:compatibility}
\sqrt{X_t(K_1)[0] - \sum_{j=2}^m \left ( u^{(j)}_t({K_1},S_t,X_t(K_1))[0]\right )^2 \ln^2\left ( \frac{{K_1}}{S_t} \right )} - u_t^{(1)}({K_1},S_t,X_t(K_1))[0] \ln\left ( \frac{{K_1}}{S_t} \right )= \\
\sqrt{X_t(K_2)[0] - \sum_{j=2}^m \left ( u^{(j)}_t({K_1},S_t,X_t(K_2))[0]\right )^2 \ln^2\left ( \frac{{K_1}}{S_t} \right )} - u_t^{(1)}({K_1},S_t,X_t(K_2))[0] \ln\left ( \frac{{K_1}}{S_t} \right ).
\end{multline}
For the difficulties in trating the multi-strike case see \cite{SchweizerWissel2007-strike}.

\bibliography{biblio}
\bibliographystyle{plain}

\end{document}